\documentclass[aps,prl,letterpaper,lengthcheck,superscriptaddress,showpacs]{revtex4}  

\usepackage{graphicx}

\begin{document}

\preprint{V2}

\title{Long-Distance Atom-Photon Entanglement}

\author{W. Rosenfeld} \affiliation{Fakult\"at f\"ur Physik,
Ludwig-Maximilians-Universit\"at M\"unchen, D-80799 M\"unchen, Germany}

\author{F. Hocke} \affiliation{Fakult\"at f\"ur Physik,
Ludwig-Maximilians-Universit\"at M\"unchen, D-80799 M\"unchen, Germany}

\author{F. Henkel} \affiliation{Fakult\"at f\"ur Physik,
Ludwig-Maximilians-Universit\"at M\"unchen, D-80799 M\"unchen, Germany}

\author{M. Krug} \affiliation{Fakult\"at f\"ur Physik,
Ludwig-Maximilians-Universit\"at M\"unchen, D-80799 M\"unchen, Germany}

\author{J. Volz} \affiliation{Fakult\"at f\"ur Physik,
Ludwig-Maximilians-Universit\"at M\"unchen, D-80799 M\"unchen, Germany}

\author{M. Weber} \email[corresponding author:
]{markus.weber@physik.uni-muenchen.de} \affiliation{Fakult\"at f\"ur Physik,
Ludwig-Maximilians-Universit\"at M\"unchen, D-80799 M\"unchen, Germany}

\author{H. Weinfurter} \affiliation{Fakult\"at f\"ur Physik,
Ludwig-Maximilians-Universit\"at M\"unchen, D-80799 M\"unchen, Germany}
\affiliation{Max-Planck-Institut f\"ur Quantenoptik, 85748 Garching, Germany}



\date{\today}

\begin{abstract}
We report the observation of entanglement between a single trapped atom and a
single photon at remote locations. The degree of coherence of the entangled
atom-photon pair is verified via appropriate local correlation measurements,
after communicating the photon via an optical fiber link of 300 m length. In
addition we measured the temporal evolution of the atomic density matrix after
projecting the atom via a state measurement of the photon onto several well
defined spin states. We find that the state of the single atom dephases on a
timescale of 150 $\mu$s, which represents an important step toward
long-distance quantum networking with individual neutral atoms.
\end{abstract}



\maketitle


Entanglement between light and matter
\cite{Blinov04,Matsukevich05,Volz06,Sherson06} plays an outstanding role in
long-distance quantum communication, allowing efficient distribution of
quantum information over, in principle, arbitrary large distances. By
interfacing matter-based quantum processors and photonic communication
channels, light-matter entanglement is regarded as fundamental building block
for future applications such as the quantum repeater \cite{Briegel98} and
quantum networks. In addition, this new kind of entanglement would allow,
e.g., quantum teleportation \cite{Bennett93} of quantum states of light onto
matter \cite{Sherson06,Chen08} as well as the heralded generation of
entanglement between quantum memories \cite{Moehring07} via entanglement
swapping \cite{Zukowski93}. Light-matter entanglement is thus not only crucial
for long range quantum communication but forms the basis for a first
loophole-free test of Bell's inequality with a pair of entangled atoms at
remote locations \cite{Simon03,Volz06}.

So far, three different approaches entangling light and matter have been
pursued. The spontaneous decay in a lambda-type transition of a single trapped
atom/ion enables one to entangle the internal degree of freedom of the emitted
photon with the spin-state of the atom \cite{Blinov04,Volz06}. Experiments in
this direction recently achieved the observation of entanglement between two
individually trapped ions \cite{Moehring07}, the remote preparation of an
atomic quantum memory \cite{Rosenfeld07}, and the realization of a single-atom
single-photon quantum interface based on optical high-Q cavities
\cite{Wilk07}. Other approaches are based on entanglement between
coherently scattered photons and collective spin-excitations in atomic
ensembles \cite{Matsukevich05,Chou05,Chen08}, and entanglement
between continuous variables of light and matter \cite{Julsgaard04,Sherson06}.

The relevance of light-matter entanglement for quantum networking arises from
the fact that it establishes non-classical correlations between a localized
matter-based quantum memory and an optical carrier of quantum information
which can easily be sent to a distant location. Together with appropriate
quantum communication protocols like quantum teleportation this allows to map
photonic quantum information (QI) into quantum memories, to buffer QI, and to
reconvert it later on again to photonic quantum carriers. In this context
decoherence of the photonic quantum channel as well as decoherence of the
matter-based quantum memory are important figures of merit setting a limit how
far quantum information can be distributed or how long this information can be
stored, respectively. Therefore, the ability to generate and preserve
light-matter entanglement over large distances \cite{Riedmatten06} opens the
possibility for long-distance distribution of quantum information
\cite{Simon07}.

In this Letter, we report the first direct observation of entanglement between
the internal state of a single trapped $^{87}$Rb atom and the polarization
state of a single photon which has passed 300 m optical fiber. This is
achieved by actively stabilizing both the birefringence of the optical
fiber-link as well as ambient magnetic fields in order to minimize dephasing
of the atomic memory qubit, stored in the atomic ground state 5$^2S_{1/2},
F=1, m_F=\pm1$. Detailed coherence measurements of the atomic qubit show that
photonic quantum information can be stored for about 150 $\mu$s.

\begin{figure}[t]
\includegraphics[width=7cm]{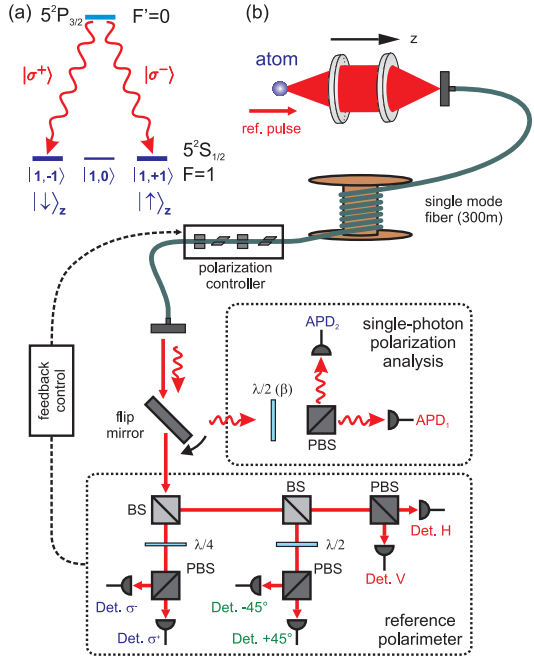}
\caption{(Color online). Schematic of long-distance atom-photon
  entanglement. (a) During the spontaneous decay of a single optically trapped
  $^{87}$Rb atom on the transition 5$^2P_{3/2}, F'=0 \rightarrow$ 5$^2S_{1/2},
  F=1$ the polarization of the emitted photon gets entangled with the final
  spin-state of the atom. (b) The emitted photon is coupled into a single-mode
  optical fiber and communicated to a remote location where a polarization
  analysis is performed. To overcome thermally and mechanically induced
  fluctuations of the fiber birefringence, an active polarization compensation
  is used. Therefore reference laser pulses are sent through the optical fiber
  and the output polarizations are characterized in a reference
  polarimeter. With the help of a software algorithm and a dynamic
  polarization controller it is ensured that input and output polarizations
  are identical.
\label{Fig1}}
\end{figure}


\begin{figure}[t]
\includegraphics[width=7cm]{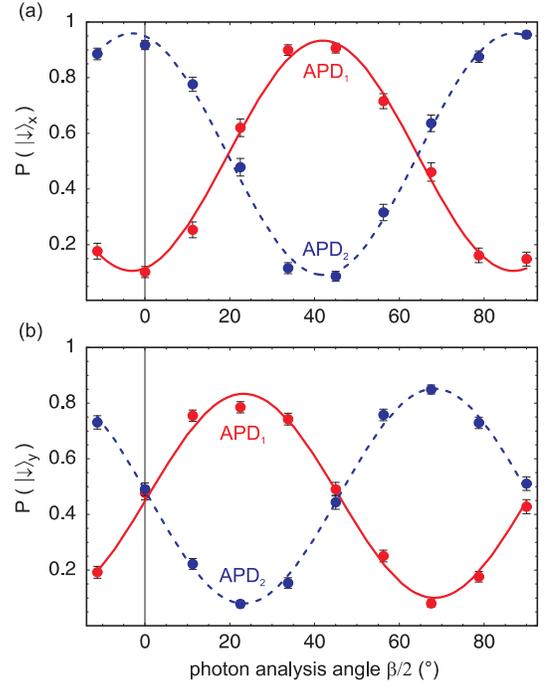}
\caption{(Color online). Verification of atom-photon entanglement. Probability
  of detecting the atomic qubit in (a)
  $\left|\downarrow\right\rangle_x:=1/\sqrt{2}(\left|\uparrow\right\rangle_z -
  \left|\downarrow\right\rangle_z)$ and (b)
  $\left|\downarrow\right\rangle_y:=1/\sqrt{2}(\left|\uparrow\right\rangle_z -
  i \left|\downarrow\right\rangle_z)$ conditioned on the detection of the
  photon in detector APD$_1$ or APD$_2$, where the photonic qubit is projected
  onto the states $1/\sqrt{2}(|\sigma^+\rangle \pm
  e^{2i\beta}|\sigma^-\rangle)$. The phase $\beta$ can be set with a rotatable
  $\lambda/2$ waveplate in front of a polarizing beam splitter (PBS). For
  $\beta=0^{\circ}$ respectively $\beta/2=22.5^{\circ}$ the photon is analyzed
  in the complementary measurement bases $\sigma_x$ and
  $\sigma_y$. \label{Fig2}}
\end{figure}

In our experiment, entanglement between the spin of a single optically trapped
$^{87}$Rb atom and the polarization of a photon is generated in the
spontaneous emission process in a lambda-type transition \cite{Volz06},
resulting in the maximally entangled atom-photon state
\begin{equation}
|\Psi\rangle = \frac{1}{\sqrt{2}} (|1,-1\rangle |\sigma^+ \rangle +
                                   |1,+1\rangle |\sigma^- \rangle ).
\end{equation} 
Here the two circular polarization states $\{\left|\sigma^{+}\right\rangle,
|\sigma^{-}\rangle\}$ of the photon define the photonic qubit, and the angular
momentum states $\{\left|F=1,m_F=-1\right\rangle:=$
$\left|\downarrow\right\rangle_z$, $\left|F=1,m_F=+1\right\rangle:=$
$\left|\uparrow\right\rangle_z\}$ the atomic qubit, respectively. While the
atom is spatially localized in the optical dipole trap, the emitted photon is
coupled into a single mode optical fiber and guided to a remote location where
a polarization analysis is performed. To measure and compensate drifts of the
fiber birefringence, reference laser pulses with two complementary
polarizations $(V,+45^{\circ})$ are sent through the optical fiber
(incorporating a fiber-based dynamic polarization controller) and the
respective output polarizations are analyzed with a reference polarimeter (see
Fig. \ref{Fig1} (b)). Based on the difference between input and output
polarization a software algorithm calculates new parameters for the dynamic
polarization controller thereby optimizing the alignment. One such step takes
0.7 s, which is currently limited by the switching speed of the
opto-mechanical shutters. These steps are repeated iteratively until input and
output polarizations are identical within $99.9\%$ \cite{Hocke07}. Once the
algorithm has compensated the fiber birefringence, typically after 10 steps,
single photons from the atom are sent through the fiber. To verify atom-photon
entanglement the internal atomic spin state is measured locally in two
complementary measurement bases $\sigma_x$ and $\sigma_y$ with the help of a
Stimulated-Raman-Adiabatic-Passage (STIRAP) technique \cite{Vewinger03,Volz06}
and correlated with the polarization analysis of the photon. Typically, the
bare photon detection efficiency is $1.2 \times 10^{-3}$, including coupling
losses into the single mode optical fiber and the limited quantum efficiency
of the single photon detectors. Together with transmission losses in the 300 m
optical fiber and coupling losses of the dynamic polarization controller this
results in the total detection efficiency of $0.6 \times 10^{-3}$. The final
event rate of $15$ min$^{-1}$ is mainly caused by frequent reloading of the
dipole trap.

For the analysis of entanglement, we determined the probability of detecting
the atomic qubit in $\left|\downarrow\right\rangle_x$ and
$\left|\downarrow\right\rangle_y$ conditioned on the projection of the photon
onto the states $1/\sqrt{2}(|\sigma^+\rangle \pm e^{2i\beta}|\sigma^-\rangle)$
(see Fig. \ref{Fig2} (a) and (b)). For $\beta=0^{\circ}$ APD$_1$ and APD$_2$
analyze the photonic qubit in the eigenstates $|H\rangle:=$
$1/\sqrt{2}(|\sigma^+\rangle + |\sigma^-\rangle)$ and $|V\rangle:=$
$1/\sqrt{2}(|\sigma^+\rangle - |\sigma^-\rangle)$ of $\hat{\sigma}_x$, whereas
for $\beta=45^{\circ}$ APD$_1$ and APD$_2$ project onto the eigenstates
$\left|+45^{\circ}\right\rangle:=$ $1/\sqrt{2}(|\sigma^+ \rangle - i
|\sigma^-\rangle)$ and $\left|-45^{\circ}\right\rangle:=$
$1/\sqrt{2}(|\sigma^+\rangle + i |\sigma^-\rangle)$ of $\hat{\sigma}_y$. As
expected, if a $|V\rangle$-polarized photon is detected, the atom is found
with high probability in the corresponding state
$\left|\downarrow\right\rangle_x$, whereas if a $|H\rangle$-polarized photon
is registered the atom is with low probability in the state
$\left|\downarrow\right\rangle_x$. Observing similar correlations in the
complementary $\sigma_y$ basis of the atom (see Fig. \ref{Fig2}(b)) confirms
the entanglement. The measurements in Fig. \ref{Fig2} show that the observed
atom-photon pair is in the entangled state $|\Psi\rangle$ (see Eq. 1). To
determine the degree of entanglement, sinusoidal functions were fitted onto
the measured atom-photon correlation data. From the fits we infer a visibility
of $V_{\sigma_x}=0.85\pm0.03$ for the analysis of the atomic qubit in
$\sigma_x$ and $V_{\sigma_y}=0.75\pm0.03$ for $\sigma_y$, respectively. The
limited visibility of the atom-photon correlations is caused mainly by errors
in the atomic state detection ($7\%$), accidental photon detection events due
to dark counts of the single photon detectors ($3\%$), errors in the
preparation of the initial state ($1\%$), polarization drifts in the optical
fiber during successive stabilization sequences of the dynamic polarization
compensation ($1\%$), and residual shot-to-shot dephasing of the atomic qubit
due to fluctuations of the ambient magnetic field. For the significantly
reduced visibility in the atomic $\sigma_y$ basis compared to $\sigma_x$ we
suppose residual magnetic fields along the $x$ axis, which lead to Larmor
precession into the additional Zeeman sublevel $m_F=0$ of the 5$^2S_{1/2},
F=1$ hyperfine ground level \cite{Rosenfeld08a}. To estimate the atom-photon
entanglement fidelity $F_{at-ph}$ we assume that errors in the atomic and
photonic state detection are isotropic in all three complementary measurement
bases (white noise). Herewith we derive a minimum fidelity $F_{at-ph}$ of
$0.85\pm0.02$.


\begin{figure}[t]
\includegraphics[width=7cm]{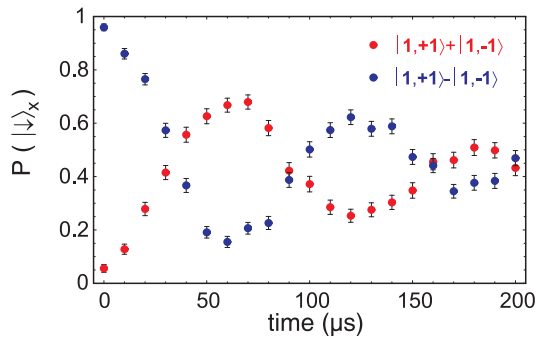}
\caption{(color online). Spin-precession of the atomic states
  $\{\left|\uparrow\right\rangle_x:=$$1/\sqrt{2}(\left|\uparrow\right\rangle_z+\left|\downarrow\right\rangle_z)$,
  $\left|\downarrow\right\rangle_x:=$$1/\sqrt{2}(\left|\uparrow\right\rangle_z-
  \left|\downarrow\right\rangle_z) \}$, as a guiding field of $5.5$ mG is
  applied along the quantization axis $z$. After the temporal evolution the
  population $P$ of the $\left|\downarrow\right\rangle_x$ state is
  measured. \label{Fig3}}
\end{figure}

For future applications in long-distance quantum communication, absorption
losses and depolarization of the photonic qubit in optical fibers are
important figures of merit \cite{Huebel07}. However, not only decoherence
effects of the photon limit the distance over which quantum information can be
distributed. Main criterion for these applications is the ability to store
quantum information in quantum memories, which show long coherence times. So
far, for quantum memories based on Zeeman qubits of neutral atoms,
experimental coherence times of several 10 $\mu$s have been demonstrated
\cite{Riedmatten06,Chen08}. In principle, clock-state quantum memories can
have much longer coherence times \cite{Kuhr03}, yet, manipulation of the
corresponding light-matter entanglement is far less practical. In our case,
fluctuating magnetic fields play an important role as they lead to
decoherence/dephasing of the atomic Zeeman qubit $|1,\pm 1\rangle$ and
consequently to decoherence/dephasing of the entangled atom-photon state.

In order to carefully distinguish between photonic and atomic decoherence we
characterized the coherence properties of the atomic quantum memory by
measuring the precession of the atomic spin in a magnetic field. This is
achieved via quantum state tomography of the atomic ground level 5$^2S_{1/2},
F=1$, reconstruction of the respective density matrix
$\rho=r|\chi\rangle\langle\chi| +(1-r)\hat{1}/3$ \cite{Rosenfeld08a}, and
determination of the corresponding purity parameter $r$. In contrast to the
fidelity $F=\left\langle \Phi|\rho|\Phi\right\rangle$ which is the overlap
between the measured density matrix $\rho$ and a pure target state
$\left|\Phi\right\rangle$, the purity parameter, here $r=\sqrt{1/2(3
  tr(\rho^2) -1)}$, is related to the coherent fraction of the density matrix
with respect to the \textit{closest} pure state $|\chi\rangle$ (which is in
general unknown). Therefore $r$ is ideally suited to quantify decoherence
effects of our atomic quantum memory.

In our spin-precession experiments, the 300 m optical fiber of the first
experiment is replaced by a 5 m one, leading to a negligible time delay of
$25$ ns between the preparation of the entangled atom-photon pair and the
initialization of the atomic spin-state via a projective polarization
measurement of the photon. The magnetic bias field is controlled via
additional Helmholtz coils and an active feedback loop with an accuracy of
$|B|<2$ mG. After the atomic spin has freely evolved for defined time periods
in the magnetic field, tomography of the final atomic spin state was performed
by measuring populations of the atomic eigenstates of the Pauli spin-operators
$\hat{\sigma}_x$, $\hat{\sigma}_y$, and $\hat{\sigma}_z$
\cite{Rosenfeld07}. In the case where a small magnetic guiding field of $5.5$
mG is applied along the quantization axis $z$ we observe the expected Larmor
precession of a spin-1/2 atom (see Fig. \ref{Fig3}), with a 1/e dephasing time
of 150 $\mu$s. In the general case where magnetic guiding field is not along
the z-axis or in the case where no guiding field is applied, the atom can
precess out of the qubit subspace $\{|1,-1\rangle, |1,+1\rangle\}$ into the
$|F=1,m_F=0\rangle$ Zeeman state of the $5^2S_{1/2}, F=1$ hyperfine ground
level. Thus, for complete characterization of decoherence effects it is
necessary to reconstruct the $3\times3$ spin-1 density matrix $\rho$. This is
possible with certain constraints. Coherences between the states
$\left|1,\pm1\right\rangle$ and $\left|1,0\right\rangle$ can not be measured
with the present atomic state detection technique, as the applied STIRAP
pulses analyze only the $\{\left|1,-1\right\rangle, \left|1,+1\right\rangle\}$
qubit subspace in a complete way \cite{Volz06,Rosenfeld07}. However, the
population in the $|1,0\rangle$ state can be inferred as the population
missing in the $\{|1,-1\rangle, |1,+1\rangle\}$ subspace. To reconstruct the
density matrix $\rho$ of the spin-1 state, we apply a worst-case assumption
that there is no coherence between the $|1,0\rangle$ state and the others, and
set the corresponding components to $0$. The resulting purity parameter $r$ thus
is a conservative lower bound on the effective coherence of the atomic state.

\begin{figure}[t]
\includegraphics[width=7cm]{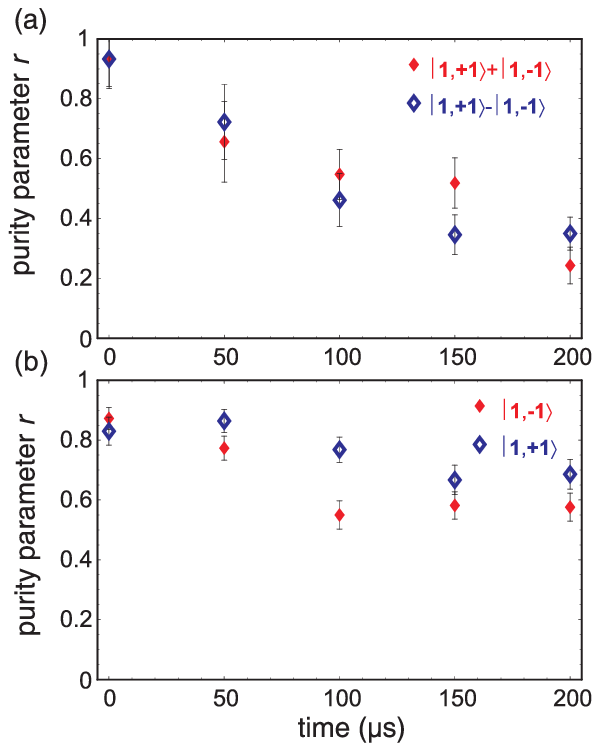}
\caption{(color online). Temporal evolution of the lower bound of the purity
  parameter $r$ as the atom is prepared initially in the states (a)
  $\{\left|\uparrow\right\rangle_x:=$$1/\sqrt{2}(\left|\uparrow\right\rangle_z+\left|\downarrow\right\rangle_z)$,
  $\left|\downarrow\right\rangle_x:=$$1/\sqrt{2}(\left|\uparrow\right\rangle_z-
  \left|\downarrow\right\rangle_z) \}$, (b) $\{
  \left|\uparrow\right\rangle_z:=\left|1,+1\right\rangle,$
  $\left|\downarrow\right\rangle_z:=\left|1,-1\right\rangle \}$. Each
  measurement point results from a partial quantum state tomography of the
  spin-1 ground level $5^2S_{1/2}, F=1$. \label{Fig4}}
\end{figure}

In a second measurement run no guiding field is applied (corresponding to the
situation of a magnetic zero field with an accuracy of $\pm 2$ mG) and
dephasing of the atomic superposition states
$1/\sqrt{2}(|1,-1\rangle\pm|1,+1\rangle)$ is analyzed by reconstruction of the
minimal purity parameter $r$. Here we find transversal $1/e$ dephasing times
of $T_2^* = 75..150$ $\mu$s, see Fig. \ref{Fig4}(a). For the states $|1,\pm
1\rangle$, see Fig. \ref{Fig4}(b), the longitudinal dephasing times are
estimated by extrapolation to be $> 0.5$ ms. The faster dephasing of
superposition states shows, that fluctuations respectively shot-to-shot noise
of the effective magnetic field are mainly along the quantization axis
$z$. This effect is due to a small fraction of circularly polarized
dipole-trap light (below 1 $\%$), which leads in combination with a finite
atomic temperature of 150 $\mu$K to a position-dependent differential
light-shift \cite{Rosenfeld08a}. 

In this Letter, we successfully demonstrated the generation and verification
of entanglement between a single trapped neutral atom and single photon
separated by 300 m optical fiber. Our implementation includes an active
stabilization of ambient magnetic fields with an accuracy of $|B|<2$ mG,
resulting in a dephasing time of the atomic memory level $5^2S_{1/2}, F=1$ of
$\simeq 150$~$\mu$~s. Longer coherence times could be reached with higher
accuracy of the polarization of the dipole trap light, lower temperature of
the trapped atom, and better stability of the magnetic field. Nevertheless,
together with the implemented stable optical fiber link also the current setup
should allow to entangle two optically trapped $^{87}$Rb atoms at locations
spatially separated by several 100 m, ready for future applications in
long-distance quantum networking with neutral atoms and a loophole-free test
of Bell's inequality \cite{Volz06}.

\begin{acknowledgements}
This work was supported by the Deutsche Forschungsgemeinschaft, the European
Commission through the EU Project QAP (IST-3-015848), the Elite Network of
Bavaria through the excellence program QCCC, and the Munich-Centre for
Advanced Photonics MAP.
\end{acknowledgements}

\end{document}